\documentclass[aps, prb ,twocolumn, showpacs, showkeys, amssymb, amsmath]{revtex4}

\newcommand{\vo}{\mbox{VO$_2$}}

\newcommand{\vooo}{\mbox{V$_2$O$_5$}}
\usepackage{graphicx}
\usepackage{float}
\usepackage[dvips]{epsfig}
\usepackage{bm}
\usepackage{amsmath}
\usepackage{dcolumn}

\begin{document}

\title{Mott-Hubbard insulator-metal transition in the $\vo$ thin film: A combined XAS and resonant PES study}
\author{S.S. Majid$^{1}$, K. Gautam$^{2}$, A. Ahad$^{1}$, F. Rahman$^{1}$, R.J. Choudhary$^{2}$, Frank M. F. De Groot$^{3}$, D.K. Shukla$^{2}$} 
\email{dkshukla@csr.res.in}
\affiliation{$^1$ Department of Physics, Aligarh Muslim University, Aligarh-202002, India\\ 
$^2$ UGC-DAE Consortium for Scientific Research, Indore-452001, India\\
$^3$ Debye Institute for Nanomaterials Science, Utrecht University-99, 3584 CG, Utrecht, The Netherlands}

\date{\today}

\begin{abstract}
We have analyzed spectral weight changes in the conduction and the valence band across insulator to metal transition (IMT) in the $\vo$ thin film using X-ray absorption spectroscopy (XAS) and resonant photoemission spectroscopy (PES). Through temperature dependent XAS and resonant PES measurements we unveil that spectral changes in the d$_{\|}$ states (V 3$\it{d_{x^2-y^2}}$ orbitals) are directly associated with temperature dependent electrical conductivity. Due to presence of the strong electron-electron correlations among the d$_{\|}$ states, across IMT, these states are found to exhibit significant intensity variation compared to insignificant changes in the $\pi^{\ast}$ and the $\sigma^{\ast}$ states (which are O 2$\it{p}$ hybridized V 3$\it{d}$ $e_g^{\pi}$ and $e_g^{\sigma}$ states) in the conduction band. Experimentally obtained values of the correlation parameter (U$_{dd}$ $\sim$ 5.1 eV, intra-atomic V 3$\it{d}$ correlations) and crystal field splitting (10 Dq $\sim$ 2.5 eV) values are used to simulate the V $\it{L_{2,3}}$ edge XAS spectra and an agreement between simulated and experimental spectra also manifests strong correlations. These results unravel that the IMT observed in the $\vo$ thin film is the Mott-Hubbard insulator-metal transition.  
\end{abstract}

\maketitle

\section{Introduction}
Vanadium dioxide ($\vo$) is an intriguing oxide material which undergoes first order transition from low temperature insulating monoclinic $\it{(P2_1/c)}$ structure to high temperature metallic rutile structure $\it{(P4_2/mnm)}$ at T$_t$ $\sim$ 341 K $\cite{Zyl75,Ima98}$. $\vo$ has attracted attention of researchers from last several decades because of illusive nature of the IMT and possible potential applications such as thermochromic windows, memory devices and ultrafast optical switches~\cite{Pan17,Liu15,Nak12,Qaz07,Cav01}. The driving mechanism of the IMT in the $\vo$ is under debate, whether structural driven, involving electron-phonon interactions (Peierls model), or associated with the electron-electron correlations (Mott-Hubbard model). Both, theoretically and experimentally significant efforts have been made to completely understand the physics of the IMT, but a comprehensive understanding about relative roles of the Mott-Hubbard and the Peierls mechanism is still missing. Biermann $\it{et}$$\it{al}$.,$\cite{Bie05}$ concluded correlation-assisted Peierls transition while Haverkort $\it{et}$$\it{al}$.,$\cite{Hav05}$ have shown orbital-assisted collaborative Mott-Peierls transition. The insulator to metal transition in the $\vo$ accompany concomitant spectral changes in the valence and the conduction band and are relevant for addressing the questions regarding origin of the IMT $\cite{Aet13,Qua17,Lav12}$. In this context study of the electronic structure and role of the orbital occupation around the Fermi level (E$_F$) across the IMT in the $\vo$ becomes important.
 
\par The electronic structure of the $\vo$ can be understood within a crystal field model using simple molecular orbital theory $\cite{Goo73}$. In the rutile metallic phase of the $\vo$, octahedral crystal field splits the V 3$\it{d}$ degenerate orbitals into the doubly degenerate $e_g^{\sigma}$ and the triply degenerate $t_{2g}$ orbitals. The small orthorhombic component of the crystal field associated with different equatorial and apical V$\textendash$O distances further splits the triply degenerate $t_{2g}$ orbitals into the doubly degenerate $e_g^{\pi}$ orbitals and a single $\it{d_{x^2-y^2}}$ (non-bonding d$_{\|}$) orbital. The $e_g^{\pi}$ orbitals hybridize with the O 2$\it{p}$ orbitals to form bonding ($\pi$) and anti-bonding ($\pi^{\ast}$) molecular orbitals, while the d$_{\|}$ orbital is oriented along the $\it{c}$ axis. We have employed XAS and resonant PES techniques, both, to investigate the electronic band structure of the unoccupied (conduction band) and the occupied (valence band) states, respectively. XAS is an invaluable tool to study the conduction band while the resonant PES reveals the true nature of the valence bands $\cite{Gro01,Smi88}$.
   
\par We have carried out temperature dependent XAS and resonant PES measurements on the $\vo$ thin film, grown on Silicon substrate, to unveil spectral changes in the orbital occupations and their role in the IMT. Details of synthesis, structural and electrical studies of the $\vo$ thin film used in this study are reported elsewhere $\cite{Maj18}$. Significant changes in spectral weight of the one dimensional (1D) charactered d$_{\|}$ orbitals is observed across the IMT and is crucial for understanding the IMT in the $\vo$ thin film. The V $\it{L_{2,3}}$ edge XAS spectra have been simulated to match the experimental XAS spectra in the insulating and the metallic phases of the $\vo$ using charge transfer multiplet (CTM) calculations. Findings in this study confirm presence of the strong electron-electron correlations in the $\vo$ thin film.

\section{Experimental details}
High quality [0 0 1] oriented monoclinic phase $\vo$ thin film has been grown on cheap and commercially available Silicon substrate using pulsed laser deposition technique $\cite{Maj18}$. Temperature dependent soft X-ray absorption spectroscopy (SXAS across the V $\it{L_{2,3}}$ and the O $\it{K}$ edges were carried out in total electron yield (TEY) mode at the beam line BL-01, Indus-2 at RRCAT, Indore, India. Energy resolution during SXAS measurements at the oxygen $\it{K}$ edge energy was $\sim$ 250 meV. Bulk $\vooo$ was measured for reference. Photoemission spectroscopy were performed at the angle-integrated photoemission spectroscopy (AIPES) beamline, Indus-1 synchrotron radiation source at RRCAT, Indore, India. In order to study resonance effects in the photoemission process, valence band spectra (VBS) were collected (at 300 K and 373 K) with incident photon energies varying from 40 eV to 60 eV, to cover the V 3$\it{p}$ $\rightarrow$ 3$\it{d}$ transition. The sample surface was cleaned by 500 eV Ar$^+$ ions before measuring the valence band spectra. The Fermi level was aligned by recording the valence band spectra of the Ag foil, which was mounted along with the sample. Experimental resolution during the VBS measurements was estimated to be $\sim$ 0.25 $\textendash$ 0.27 eV. 

\section{Results}
\begin{figure}
\centering
\includegraphics[width=0.5\textwidth]{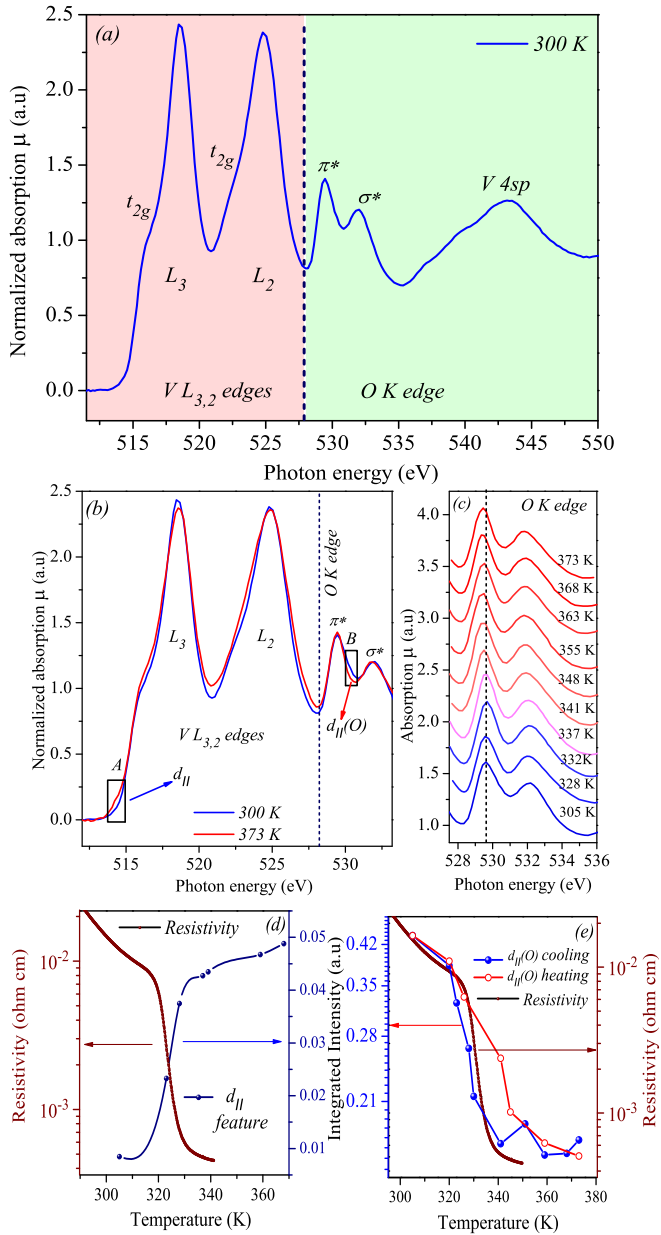} 
\caption{(a) The room temperature V $\it{L_{2,3}}$ and O $\it{K}$ edge XAS spectrum of the $\vo$ thin film. (b) Comparison of near edge XAS spectra of the insulating (300 K) and the metallic (373 K) phases. (c) The temperature dependent O $\it{K}$ near edge XAS spectra of the $\vo$ thin film. (d,e) Temperature dependence of spectral weight of the features A (d$_{\|}$) and B (d$_{\|}$ (O)) along with resistivity curve of the $\vo$ thin film. All temperature dependent data shown in this figure are from the cooling cycle except data of the feature B which is shown in both the heating and the cooling cycles to represent the hysteresis behavior of the first order phase transition.}
\label{T_XAS}
\end{figure}

Figure ~\ref{T_XAS} (a) shows room temperature V $\it{L_{2,3}}$ and O $\it{K}$ edge X-ray absorption spectrum of the $\vo$ thin film. The V $\it{L_{2,3}}$ spectrum arises due to transition from the V 2$\it{p}$ core level (spin orbit splitted V 2$\it{p_{3/2}}$ and 2$\it{p_{1/2}}$) to the unoccupied V 3$\it{d}$ states and consists of two pronounced features at $\sim$ 518.8 eV (L$_3$) and $\sim$ 525.5 eV (L$_2$). Features at $\sim$ 516.2 eV and $\sim$ 522.2 eV are due to transitions to crystal field splitted t$_{2g}$ part of the V 3$\it{d}$ band $\cite{Hav05,Zaa85}$. The O $\it{K}$ edge spectrum results from transition of the O 1$\it{s}$ electrons to the O 2$\it{p}$ states hybridized with the V 3$\it{d}$ orbitals. Peaks at $\sim$ 529.6 eV and $\sim$ 532.1 eV are due to transitions into the $\pi^{\ast}$ and the $\sigma^{\ast}$ hybridized bands $\cite{Ruz07,Goo73}$. The broad structures at higher energies in region $\sim$ 537 eV to 550 eV are attributed to transitions into the V 4$\it{sp}$ bands hybridized with the O 2$\it{p}$ orbitals $\cite{Abb93}$. In order to understand the spectral changes in the conduction band across the IMT XAS spectra of metallic and insulating phases are overlapped as shown in the Figure ~\ref{T_XAS} (b). The pre peak feature (A) at $\sim$ 514.1 eV in the metallic phase is assigned to the d$_{\|}$ band, representing the V 3$\it{d}$$_{\|}$-V 3$\it{d}$$_{\|}$ hybridizations. The feature A almost vanishes in the insulating phase of the $\vo$ and is signature of the IMT in the $\vo$ thin film. This is in accordance to a recent report by Yeo $\it{et}$$\it{al}$., $\cite{Yeo15}$. The feature B, designated as d$_{\|}$ (O) band (d$_{\|}$ (O), because of the d$_{\|}$ signature from the O $\it{K}$ edge part), observed at $\sim$ 530.4 eV in the insulating phase belongs to the unoccupied d$_{\|}$ band and is consistent with previous reports $\cite{Gra16,Aet13}$. The anomalous spectral weight transfer of the d$_{\|}$ states from $\sim$ 514.1 eV to $\sim$ 530.4 eV is observed across the IMT in the $\vo$ thin film. Such a large spectral weight transfer is considered as the fingerprinting of strong electronic correlations $\cite{Yeo15,Wan13,Rus08}$. During SXAS measurements the polarization vector of the incident photons was perpendicular to the surface normal of the thin film, monoclinic axis [0 1 1], which means that the photon polarization is parallel to the plane in which monoclinic $\it{a}$-axis (or rutile $\it{c}$-axis, as $\it{a_M}$ = 2$\it{c_R}$) of $\vo$ crystallites are randomly oriented but they all share the same surface normal [0 1 1] $\cite{Maj18}$. So, the spectral weight of both the features A and B is mainly due to averaged intensity of the V 3$\it{d}_{\|}$ states aligned along the monoclinic a-axis (or rutile c-axis) of randomly oriented $\vo$ crystallites sharing the same surface normal [0 1 1].

\begin{figure*} 
 \centering
\includegraphics[width= 0.8\textwidth]{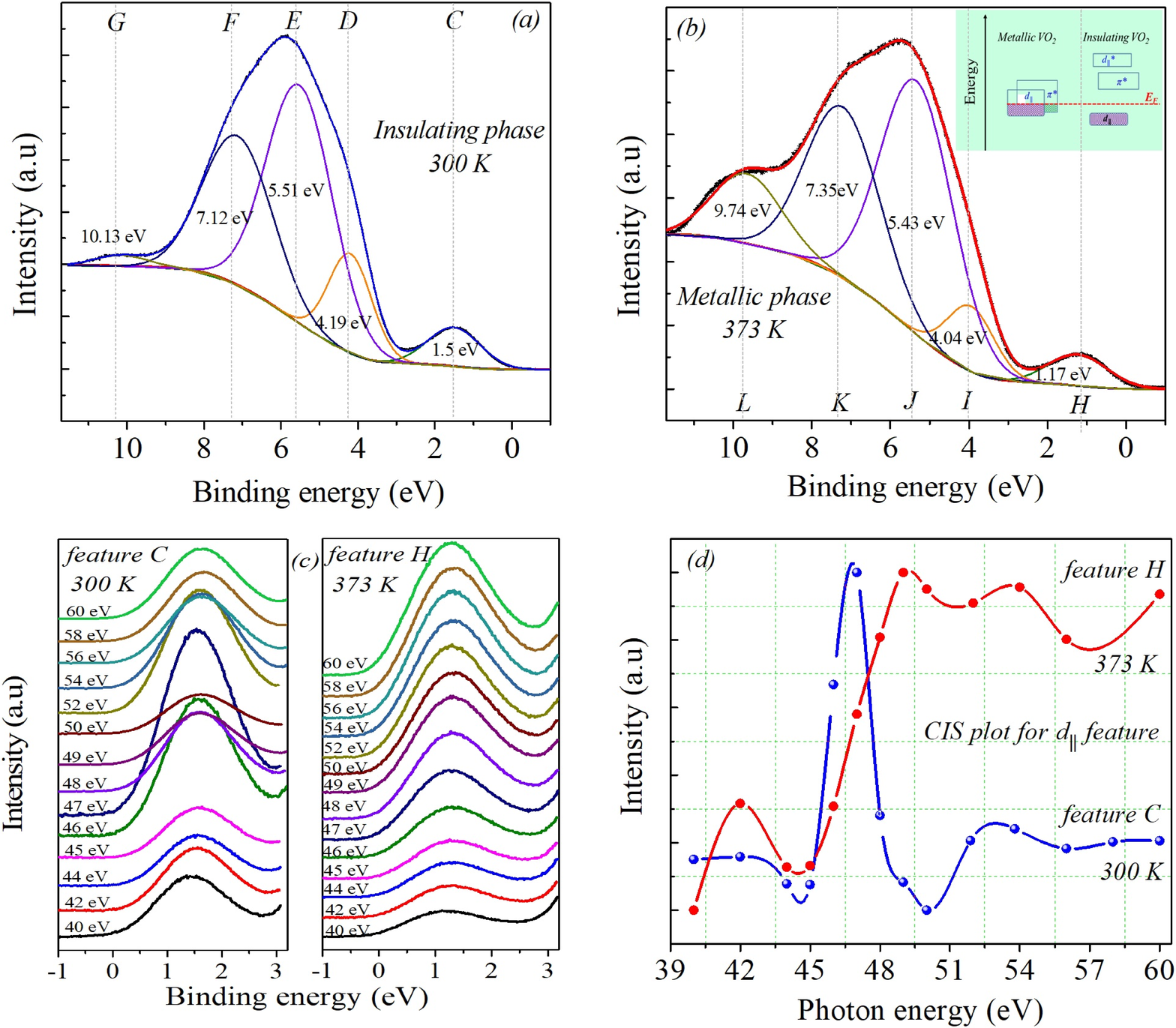} 
\caption{(a) Valence band spectrum (VBS) of the $\vo$ thin film collected using 40 eV incident photon energy in the insulating phase, 300 K (a), and the metallic phase, 373 K (b). Inset (b) shows schematic representation of the energy bands in the insulating and the metallic phases of the $\vo$ thin film. (c) Vanadium 3$\it{d}$ VBS of the $\vo$ thin film in the insulating (feature C) and the metallic (feature H) phases collected at various photon energies. (d) Constant initial state (CIS) spectra of the V 3$\it{d}$ bands (feature C in the insulating and feature H in the metallic phase).}
\label{RPES}
\end{figure*}

In the metallic phase spectral weight of the feature B vanishes while negligible spectral weight changes are observed in the $\pi^{\ast}$ and the $\sigma^{\ast}$ bands compared to the insulating phase of the $\vo$ thin film. Observation of the hysteresis in spectral weight change of the d$_{\|}$ (O) states across the IMT confirms the first-order nature of the electronic structure transition, see Figure ~\ref{T_XAS}(e). Figure ~\ref{T_XAS}(c) shows temperature dependent behavior of the O $\it{K}$ edge XAS spectra. Dotted line indicates transfer in spectral weight of the O $\it{K}$ edge towards lower photon energy with increase in temperature. With respect to room temperature insulating phase the O $\it{K}$ edge spectrum in high temperature metallic phase shifts towards lower photon energy as well as shrinks.

\par Room temperature valence band (insulating phase, 300 K) spectrum of the $\vo$ thin film measured with 40 eV photon energy is shown in the Figure ~\ref{RPES} (a). The spectrum is constituted of mainly the V 3$\it{d}$ band ranging from E$_F$ to $\sim$ 3 eV binding energy and the O 2$\it{p}$ bands above 3 eV $\cite{Gup76,Smi88,Shi90}$. The V 3$\it{d}$ region is fitted with single peak C at $\sim$ 1.5 eV while the O 2$\it{p}$ region is fitted with the peaks D, E, F and G at $\sim$ 4.19 eV, $\sim$ 7.12 eV, $\sim$ 5.53 eV and $\sim$ 10.13 eV, respectively. The peak C is assigned to the occupied V 3$\it{d}$$_{\|}$ band. Other features D to G correspond to the V 3$\it{d}$$\textendash$O 2$\it{p}$ bonding orbitals$\cite{Shi90}$. Here, we will focus mainly on the purely V 3$\it{d}$$_{\|}$ band, feature C. Figure ~\ref{RPES} (b) shows valence band spectra of the $\vo$ thin film in the metallic phase (373 K). The V 3$\it{d}$ band is fitted with a peak, H, at $\sim$ 1.2 eV and rest of the band contains four fitted features similar to the insulating phase. In order to extract true nature of the V 3$\it{d}$ valence bands above and below the IMT, measurements have been performed by recording energy distribution curves (EDCs) for different incident photon energies, ranging from 40 eV to 60 eV in both the insulating (300 K)and the metallic phase (373 K), shown in the Figure ~\ref{RPES} (c). Selected range of photon energies cover resonance of the V$^{4+}$. With variation in incident photon energy changes in intensities and spectral shape of EDCs can be readily observed, see Figure ~\ref{RPES} (c). From resonant PES measurements, one can plot the constant initial state (CIS) spectrum. A CIS spectrum from EDCs can be obtained by plotting normalized intensity of a particular feature versus incident photon energies. 

\par Resonant photoemission from the V 3$\it{d}$ atomic states occur when energy of incident photon coincides with the V 3$\it{p}$ $\rightarrow$ 3$\it{d}$ absorption threshold $\cite{Bar85,Ber83,Smi88}$. At resonant energies different ionization mechanisms (direct and indirect) leave the system in same final state. Direct photoemission from the V 3$\it{d}$ band can be written as:\par 
3$\textit{p}$$^6$3$\textit{d}$$^1$ + $\textsl{h}$$\nu$ $\rightarrow$ 3$\textit{p}$$^6$3$\textit{d}$$^0$ + e$^-$($\epsilon$),\par 
where a photon of energy $\textsl{h}$$\nu$ ionizes the V atom and an electron with kinetic energy $\epsilon$ is emitted. Indirect process involves an intra-atomic excitation process:\par
3$\textit{p}$$^6$3$\textit{d}$$^1$ + $\textsl{h}$$\nu$ $\rightarrow$ [3$\textit{p}$$^5$3$\textit{d}$$^2$]$^*$, followed by emission of a 3$\it{d}$ electron through the super Coster Kronig decay,\par
[3$\textit{p}$$^5$3$\textit{d}$$^2$]$^*$$\rightarrow$3$\textit{p}$$^6$3$\textit{d}$$^0$ + e$^-$($\epsilon$).
\par Here final state and electron kinetic energy are same as obtained in the direct photoemission process. In this case a resonance profile is observed in photoelectron intensity versus incident photon energy.
 
\par  The CIS plots for the V 3$\it{d}$ bands, features C (insulating phase) and H (metallic phase) are shown in the Figure~\ref{RPES} (d). The resonance profile of the feature C, in the insulating phase has onset of resonance started around $\sim$ 44 eV and maximizes around $\sim$ 47 eV, which is consistent with the earlier reports on the single crystal $\vo$ $\cite{Shi90}$. The CIS curve in the metallic phase after a steep rise stays nearly constant for the measured range of photon energies. These observations are consistent with changes observed in band structure in the $\vo$ across the IMT, $\cite{Goo71}$ as shown in the inset of the Figure ~\ref{RPES} (b) and confirm  that V 3$\it{d}$ peak, H, in the metallic phase is constituted of the d$_{\|}$ and the $\pi^{\ast}$ bands (inseparable due to resolution limit). In the insulating phase of the $\vo$ thin film, resonance observed in the V 3$\it{d}$ band is due to transition of the V 3$\it{p}$ electrons to the completely electron occupied d$_{\|}$ band while in the metallic phase resonance occur due to  transition of the V 3$\it{p}$ to the overlapped partially unoccupied d$_{\|}$ and $\pi^{\ast}$ bands (see inset of the Figure ~\ref{RPES}(b)). Unoccupied states allows resonance to happen, at energies even above 44 eV (resonance energy of V$^{4+}$) as there are enough available vacant states in the valence band for the V 3$\it{p}$ $\rightarrow$ 3$\it{d}$ transition to happen, at higher photon energies in the metallic phase.

\section{Discussion}
\begin{figure}
\centering
\includegraphics[width=0.45\textwidth]{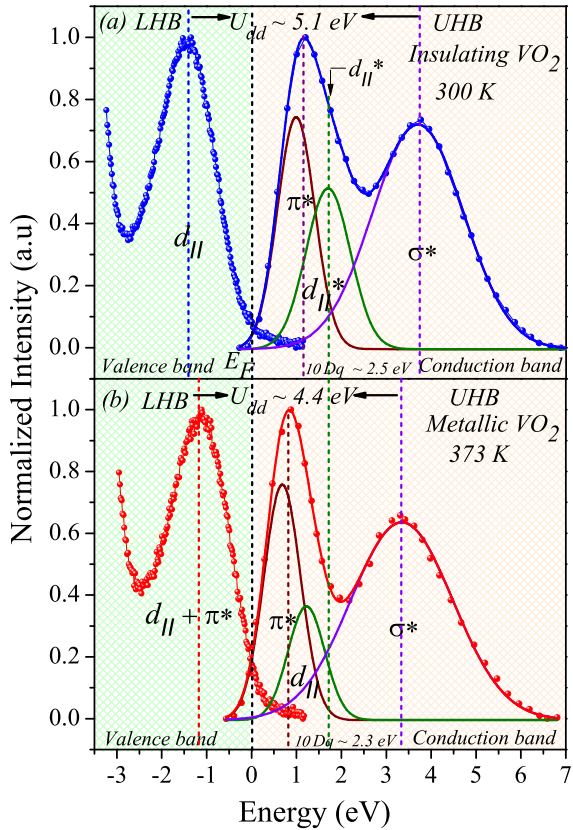} 
\caption{Approximated valence and conduction bands of the $\vo$ thin film (a) in the insulating phase and (b) in the metallic phase.} 
\label{VCB_RT_HT}
\end{figure}

Contribution of the V 3$\it{d}$ bands across the E$_F$ in valence as well as conduction band are visualized by combining the V 3$\it{d}$ bands from the VBS and the O K edge near edge XAS spectrum, see Figure ~\ref{VCB_RT_HT}. Usually, PES and the inverse PES are combined for description of the valence and the conduction band because XAS has involvement of the core hole effect. However, effect of the core hole is small in case of the oxygen K edge so it can also be used to represent the conduction band. 
Fermi level marked in Figure ~\ref{VCB_RT_HT} for the VBS is obtained with the help of Ag reference. Second derivative of the O K edge XAS was performed to determine the correct absorption onset position in XAS spectra. And to align the XAS spectra on the same axis absorption onset is taken at E$_F$ $\cite{Kur08,Koe06}$.

In valence band part, presence of finite electron density at the Fermi level in the metallic phase compared to the insulating phase indicates an electronic phase transition in the $\vo$ thin film. The valence band in the insulating phase is constituted of only the d$_{\|}$ states while in the metallic phase the d$_{\|}$ and the $\pi^{\ast}$ states both contribute $\cite{Goo71}$. The conduction band in the insulating phase (Figure ~\ref{VCB_RT_HT} (a)) is comprised of three features the $\pi^{\ast}$, the d$_{\|}^{\ast}$ and the $\sigma^{\ast}$ states. The conduction band in the metallic phase (Figure ~\ref{VCB_RT_HT} (b)) is comprised of the $\pi^{\ast}$ and the $\sigma^{\ast}$ states. In the metallic phase we observe disappearance of the d$_{\|}^{\ast}$ orbitals (which is marked with an arrow in the insulating phase). The d$_{\|}$ feature shown in the conduction band of the metallic phase is from unoccupied part of the non-bonding d$_{\|}$ states overlapping with the $\pi^{\ast}$ state. Observation of orbital switching of the d$_{\|}^{\ast}$ states across the IMT in the $\vo$ thin film is consistent with previous X-ray absorption studies on the $\vo$ single crystals $\cite{Hav05,Koe06}$.

\par The behavior of temperature dependent integrated intensity of the d$_{\|}$ states (features A and B) from XAS measurements are compared with resistivity curve, see the Figures ~\ref{T_XAS}(d,e). This clearly indicates that conductivity in the $\vo$ thin film is directly associated with spectral weight changes of the d$_{\|}$ states. Such a large and concomitant change in spectral weight of the d$_{\|}$ states across the IMT is possible only when the system is strongly correlated \cite{Koe06,Yeo15,Hav05,Abb91}. Although spectral weight change of the d$_{\|}$ states across the IMT is also related to the V-V dimerization along the rutile $\it{c}$ axis, but significant spectral change of the d$_{\|}$ states compared to almost negligible changes observed in the $\pi^{\ast}$  and the $\sigma^{\ast}$ states reflect strong electron-electron correlations present in the $\vo$ thin film, as the $\pi^{\ast}$  and the $\sigma^{\ast}$ states are also susceptible to the structural changes \cite{Ruz07,Ruz08,Koe06,Qua16,Gra16} see Figure ~\ref{VCB_RT_HT}.
\par Further, separation between the d$_{\|}$ bonding and the anti-bonding bands estimated from the energy positions of the d$_{\|}$ bonding states in the valence band and the d$_{\|}$ anti-bonding states in the conduction band, comes out to be $\sim$ 3.0 eV. According to pure one electron approximation this energy separation is equal to the 2$\it{t}$, where $\it{t}$ is the intradimer hopping integral $\cite{Yeo15}$. Local density approximation (LDA) estimates value of the $\it{t}$ as $\sim$ 0.7 eV and energy separation $\sim$ 1.4 eV, which is much smaller than the one estimated here ($\sim$ 3.0 eV) $\cite{Bie05}$. The difference of $\sim$ 1.6 eV between the calculated and the experimental values is due to presence of strong electron correlations in the $\vo$ thin film, as electron-electron correlations are not considered in the LDA.

\begin{figure}	
\centering
\includegraphics[width=0.5\textwidth]{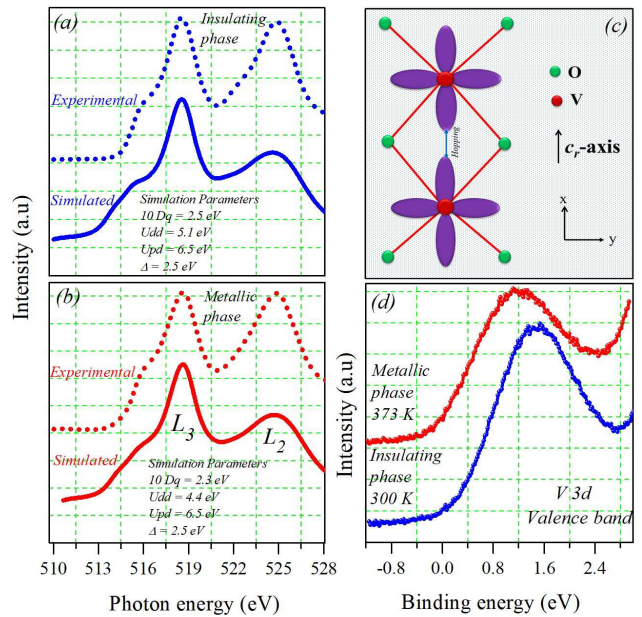} 
\caption{ The experimental and simulated V $\it{L_{2,3}}$ edge spectra (a) in the insulating and (b) in the metallic phase of the $\vo$ thin film. (c) Schematic illustration of the V 3$\it{d}$$_{\|}$ ($\it{d_{x^2-y^2}}$) and the O 2$\it{p}$ orbitals oriented along the rutile $\it{c}$ axis. (d) The V 3$\it{d}$ valence band spectra of the insulating and the metallic phases of the $\vo$ thin film.} 
\label{SXAS}
\end{figure}  

\par In order to further substantiate claim of presence of the strong electron-electron correlations in the $\vo$ thin film, we have simulated the V $\it{L_{2,3}}$ edges XAS spectra of the insulating (300 K) and the metallic (371 K) phases using relevant values obtained from the experiment, see the Figure ~\ref{SXAS} (a,b). Calculations are performed using the Charge Transfer Multiplet Model for X-ray Absorption Spectroscopy (CTM4XAS) code $\cite{Sta10}$. The CTM4XAS code is based on three components of theoretical calculations, the atomic multiplet, the crystal field and the charge transfer calculations \cite{Sta10,Hav05}. Crystal field splitting parameter, the 10 Dq, is energy between the splitted t$_{2g}$ and e$_g$ states in octahedral crystal field. The charge transfer calculations include the intra-atomic 3$\it{d}$-3$\it{d}$ Coulomb repulsion energy, U$_{dd}$, which is equal to energy required to transfer an electron from one vanadium site to another vanadium site, the core hole potential, U$_{pd}$, identified as an attractive V 2$\it{p}$3$\it{d}$ correlation energy and the charge transfer energy, $\Delta$ defined as energy difference between the V 3$\it{d}$$^1$ and the 3$\it{d}$$^2$$\underline L$ configurations, where $\underline L$ is a hole state in the O 2$\it{p}$ ligand band. The V $\it{L_{2,3}}$ edges XAS spectra is calculated assuming the V 2$\it{p^6}$3$\it{d^1}$ ($t_{2g}^1$) ground state and the Slater integrals are scaled down to 80\% of their Hartee-Fock values $\cite{Sta10,Gro01}$. Values of the U$_{dd}$, 10 Dq, U$_{pd}$ and $\Delta$ used in calculations are $\sim$ 5.1 eV, $\sim$ 2.5 eV, $\sim$ 6.5 eV and $\sim$ 2.5 eV for the insulating phase and $\sim$ 4.4 eV, $\sim$ 2.3 eV, $\sim$ 6.5 eV and $\sim$ 2.5 eV for the metallic phase. 

The Coulomb repulsion energy, U$_{dd}$ is obtained from difference between energy positions of the lower Hubbard band, LHB (d$_{\|}$ peak in the valence band) and the upper Hubbard band, UHB ($\sigma^{\ast}$ peak in the conduction band) $\cite{Maj17,Shi15}$, while the crystal field splitting energy, 10 Dq is obtained from difference in energy positions of the $\pi^{\ast}$ and the $\sigma^{\ast}$ states in the conduction band (Figure ~\ref{VCB_RT_HT}). Calculated spectra are convoluted with a Gaussian broadening of $\sim$ 0.3 eV to account for experimental resolution. Simulated spectra of both, the insulating and the metallic, phases of the $\vo$ thin film match well with their respective experimental spectra apart from lower intensity of the simulated V $\it{L_{2}}$ edge compared to the experimentally obtained $\it{L_{2}}$ edge, which is attributed to absence of experimental background. Comparatively larger broadening of the high energy $\it{L_{2}}$ edge feature is due to finite life time of the core hole leading to an increase in uncertainty in its energy, consistent with the Heisenberg$'$s principle $\cite{Hen14}$. 

\par Presence of the strong electron-electron correlations in the $\vo$ thin film indicate that electrical conductivity must be explained in light of the Mott-Hubbard model. According to which electrical conductivity can be explained in terms of competition between hopping of electrons in the V (3$\it{d}$$_{\|}$)-V (3$\it{d}$$_{\|}$) channels and their electronic correlations with the O 2$\it{p}$ and V 3$\it{d}$$_{\|}$ electrons $\cite{Yeo15}$. In the insulating phase due to high O 2$\it{p}$ participation, evident from the increased intensity of the d$_{\|}$ (O) band (Figure ~\ref{T_XAS} (e)), correlation among electrons hopping via the V (3$\it{d}$$_{\|}$)-V (3$\it{d}$$_{\|}$) channel and the O 2$\it{p}$ electrons increases. While in the metallic phase lesser O 2$\it{p}$ participation and screening due to the $\pi^{\ast}$ electrons reduces electronic correlations, so electrons can hop from the one vanadium site to another vanadium site. Presence of the $\pi^{\ast}$ electrons in the metallic phase can be seen from resonance behavior of the $\it{d}$$_{\|}$ feature (see inset Figure ~\ref{RPES} (d) and Figure ~\ref{VCB_RT_HT} (b)). Orientation of the O 2$\it{p}$ and the V 3$\it{d}$$_{\|}$ orbitals shown in the Figure ~\ref{SXAS} (c) illustrates hopping of the electrons between the vanadium sites. Implication of hopping of electrons in the metallic phase can also be directly visualized in the band width, $\it{W}$, of the vanadium valence 3$\it{d}$ band. A larger valence bandwidth in the metallic phase represents higher electron hopping compared to  the narrow valence band in the insulating phase, see Figure ~\ref{SXAS} (d).    

\section{Conclusions}
In summary, our results unravel that electrical conductivity in the $\vo$ thin film across the IMT is directly related to spectral weight change of the d$_{\|}$ states. XAS results show that intensity of the unoccupied d$_{\|}$ bands decreases with increase in conductivity of the $\vo$ thin film. Temperature dependent resonant behavior of the d$_{\|}$ states inferred from the resonant PES experiments show that spectral shape in the metallic phase contains the $\pi^{\ast}$ states (together with the d$_{\|}$ states) which screens the correlation among the vanadium dimers while in the insulating phase spectral shape is defined solely by the d$_{\|}$ states. From temperature dependent SXAS, anomalous spectral weight transfer of the d$_{\|}$ states, $\sim$ 16 eV, is observed across the IMT which directly signifies role of the strong electron-electron correlations present in the $\vo$ thin film. A large separation between the occupied and unoccupied d$_{\|}$ bands in the insulating phase ($\sim$ 3.0 eV) also vindicates the strong electron-electron correlations. Considerable value of the Coulomb repulsion energy, U$_{dd}$ ($\sim$ 5.1 eV), which is experimentally obtained and theoretically verified, further substantiates presence of the strong electron-electron correlations in the $\vo$ thin film. Finally, by combining our experimental results and simulations the IMT observed in the $\vo$ thin film is explained in context of the Mott-Hubbard model.   

\section{Acknowledgments}
Authors are thankful to Sharad Karwal and Rakesh Sah for help during XPS and XAS measurements, respectively. D.K.S. acknowledges support from DST and SERB, India, in form of an inspire faculty award (IFA-13/PH-52) and an early-career research award (ECR/2017/000712).
\bibliography{VCB}
\end{document}